# Steady Marginality: A Uniform Approach to Shapley Value for Games with Externalities

Oskar Skibski[**]

Institute of Informatics, University of Warsaw
Banacha 2, 02-097 Warsaw, Poland
`oskar.skibski@mimuw.edu.pl`

**Abstract.** The Shapley value is one of the most important solution concepts in cooperative game theory. In coalitional games without externalities, it allows to compute a unique payoff division that meets certain desirable fairness axioms. However, in many realistic applications where externalities are present, Shapley's axioms fail to indicate such a unique division. Consequently, there are many extensions of Shapley value to the environment with externalities proposed in the literature built upon additional axioms. Two important such extensions are "externality-free" value by Pham Do and Norde and value that "absorbed all externalities" by McQuillin. They are good reference points in a space of potential payoff divisions for coalitional games with externalities as they limit the space at two opposite extremes. In a recent, important publication, De Clippel and Serrano presented a marginality-based axiomatization of the value by Pham Do Norde. In this paper, we propose a dual approach to marginality which allows us to derive the value of McQuillin. Thus, we close the picture outlined by De Clippel and Serrano.

## 1 Introduction

The Shapley value is one of the most important and extensively studied solution concepts in coalitional game theory. In the environment where agents are allowed to cooperate, the Shapley value lays down a fair allocation of jointly achieved payoff. Here, fairness is built upon four axioms: (i) *Efficiency* (the whole payoff is distributed among agents); (ii) *Symmetry* (division of payoff does not depend on agents' names), (iii) *Additivity* (when two different games are combined, agent's share is equal to sum of shares in games considered independently); and (iv) *Null-player Axiom* (agent which has no impact on value of any coalition gets nothing). In his seminal work, Shapley showed that his division scheme is *unique*, i.e. no other division scheme meets all four axioms.

Since Shapley original work [1], his concept has been extended in a variety of directions. One of them are coalitional games with externalities, where a value of a coalition depends on formation of other coalitions in the system. Indeed, externalities occur in many real-life applications of coalitional games such as oligopolistic markets (where a merger is likely to affect other companies), a political scene (where the importance of a political party highly depends on the created coalitions), or a supply chain (where a large number of subcontractors increases the standardization costs). Unfortunately, they are substantially more challenging to the model than the conventional games. In particular, in the presence of externalities, the axioms proposed by Shapley are insufficient to determine a unique division of payoff. This problem was addressed in the literature by several authors who proposed more extended axiomatizations. Two important extensions of the Shapley value to games with externalities are "externality-free" value proposed by Pham Do and Norde [3] and McQuillin's value which "absorbed all externalities" [2]. Both can be considered as reference points for other extensions, as, under certain conditions, they limit the space of possible extensions at two opposite extremes.

[**] I would like to thank my supervisor Andrzej Szalas and Tomasz Michalak for their comments, suggestions and discussions.

The key role in the formula for the Shapley value is played by the marginality — an important economic concept in which evaluation of a player in a coalition is based on a difference between the coalition's values with and without the player. Specifically, the Shapley value is calculated as the weighted average of marginal contributions of players to all the coalitions in the game. This relationship of the Shapley value to the marginality was emphasized by the beautiful work of Young [8]. In Shapley's axiomatic characterization marginality comes from the Null-player Axiom which assigns zero value to every agent whose vector of marginal contributions to coalitions equals to zero. Unfortunately, as mentioned above, this axiom as well as the others do not determine the unique value in the case of games with externalities. Thus, it became a common practise in the literature, to add other, non-marginality based axioms. These axioms indeed allow for deriving the unique Shapley value extended to games with externalities, however, not in a way related to marginality as the original Shapley value did. In this context, the key additional axiom of McQuillin was recursion which required the value to be a fixed-point solution (i.e. if we consider a value to be a game by itself, then the value computed for such a game should not change). Nevertheless, marginality-based axiomatization of the McQuillin value that connects to the original Shapley value has remained unknown.

In this paper we close this gap by proposing an alternative approach to marginality. We present a new marginality axiom, which allows us to derive the extension of the Shapley value proposed by McQuillin. Our approach is dual to De Clippel and Serrano who in recent, important publication proposed a basic approach to marginality and derived "externality-free" value of Pham Do and Norde. In other words, we close the picture outlined by De Clippel and Serrano, so that the two opposite values for games with externalities that limit the space of many other extensions, are now based on the marginality principle.

Our new approach to marginality, which we call a steady marginality, differs from those proposed earlier in the literature ([5,7,6]). To compute an agent's marginal contribution to a coalition we compare the value of the coalition with the specific agent with the value of the coalition obtained by the transfer of the agent to another coalition, existing in the partition (so the number of coalitions is *steady*). We then do not include the value of a coalition in a partition when a specific agent forms its own singleton coalition.

The rest of the paper is organized as follows. In Section 2 we introduce the basic definitions and notation. In Section 3 we present our set of axioms, including our new marginality axiom which we relate to "externality-free" marginality. In Section 4 we define a new class of games and prove they form a basis of space of games with externalities. Finally, in Section 5 we prove that there exists only one value which satisfies all our axioms and that this value is equal to one proposed earlier by McQuillin. Section 6 presents different approaches to marginality in the literature. Section 7 concludes the paper and outlines future work.

## 2 Definitions and Notation

In this section we introduce the basic definitions and notation.

Let $N = \{1, 2, \ldots, n\}$ be the set of all agents. A *coalition* $S$ is any subset of agents, $S \subseteq N$. A *partition* $P$ *of* $N$ is a set of disjoint coalitions which covers the whole set of agents, i.e., $P = \{S_1, S_2, \ldots, S_k\}$ and $\bigcup_{i \in N} S_i = N$, where $S_i \cap S_j = \emptyset$ for every $i, j \in \{1, \ldots, k\}$ with $i \neq j$. The set of all partitions is denoted by $\mathcal{P}$. A coalition $S$ being a part of a partition $P$ is called an *embedded coalition* and is denoted by $(S, P)$. By $|P|$ we denote the number of coalitions in a partition $P$. The set of all embedded coalitions is denoted $EC$ and formally defined as:

$$EC \stackrel{\text{def}}{=} \{(S, P) : P \in \mathcal{P}, S \in P\}$$

As common in the literature, for $S \subset N, i \notin S, j \in S$, we define $S_{+i} \stackrel{\text{def}}{=} S \cup \{i\}$ and $S_{-j} \stackrel{\text{def}}{=} S \setminus \{j\}$. If $S \in P$ then $P_{-S} \stackrel{\text{def}}{=} P \setminus \{S\}$.

The following notation will play an important part in our paper: let $S, T \in P$ and $i \in S$. A function $\tau_i^{S,T}$ will represent the transition of an agent $i$ from $S$ to $T$:

$$\tau_i^{S,T}(P) \stackrel{\text{def}}{=} P \setminus \{S, T\} \cup \{S_{-i}, T_{+i}\}$$

In the literature it is a common convention to assume that in every partition $P \in \mathcal{P}$ an empty, artificial coalition $\emptyset \in P$ exists. In our paper we accept a different assumption that only one partition (the one with the grand-coalition) contains a special empty coalition $\emptyset$.

*Note 2.1.* For technical convenience, in our paper we use the convention that in every partition there are at least two coalitions, so we assume that in partition with only one explicitly listed coalition there also exists an empty coalition. In such a case, the grand coalition takes the form $\{N, \emptyset\}$ (and $|\{N, \emptyset\}| = 2$). However, we will not consider $(\emptyset, \{N, \emptyset\})$ as a correct embedded coalition.

The *game* (in a partition-function form) is a function $v : EC \to \mathbb{R}$ which associates a real number with every embedded coalition. For convenience, we extend the domain of $v$ and assign a zero value to every incorrect embedded coalition where $S$ is empty: $v(\emptyset, P) = 0$.

Among a collection of games we distinguish a set of *games without externalities* (or, differently, *characteristic function games*), where the value of a coalition does not depend on a partition of other agents. Formally, for each coalition $S$ and two partitions $P_1, P_2$ containing $S$ we get $v(S, P_1) = v(S, P_2)$. In this case the definition of a game can be simplified to $\hat{v} : 2^N \to \mathbb{R}$, as the only argument is a coalition $S$. Conversely, we say that the game is *with externalities* when the value for at least one coalition depends on a structure of other agents: $v(S, P_1) \neq v(S, P_2)$ for some $S$ and $P_1, P_2$ which contain $S$.

The *value* of the game is a function which assigns some payoff to every agent: $\varphi : v \to \mathbb{R}^N$. This payoff is meant to be the agent's share in the value achieved by all players united in the grand coalition: $v(N, \{N, \emptyset\})$. We are interested in a division of the payoff which is fair.

The Shapley value is defined as:

$$Sh_i(\hat{v}) = \sum_{S \subseteq N, i \in S} \frac{(|S| - 1)!(|N| - |S|)!}{|N|!} (\hat{v}(S) - \hat{v}(S_{-i}))$$

where by $\hat{v}$ we denote the game without externalities.

Shapley presented the following intuition behind his value. Assume that the agents enter the coalition in random order. Every agent $i$ brings to the set $S_{-i}$ of agents who already entered the coalition its marginal contribution $\hat{v}(S) - \hat{v}(S_{-i})$. Therefore, Shapley value of an agent $i$ is the average of all its marginal contributions for every order of the agents' arrivals.

One of the most common approaches to the extension of Shapley value for games with externalities is the *average approach* proposed by Macho-Stadler et al [4]. In this approach, from the game $v$ we create a simpler game $\hat{v}$ without externalities and define $\varphi_i(v) \stackrel{\text{def}}{=} Sh_i(\hat{v})$. The value of each coalition $S$ in $\hat{v}$ is computed as the weighted average of values of a coalition $S$ embedded in different partitions: $\hat{v}(S) = \sum_{P \in \mathcal{P}, S \in P} \alpha_{(S,P)} \cdot v(S, P)$. The different weights lead to the different values. Two extremes in those approaches are "externality-free" value and value which "absorbed all externalities".

The first one was proposed by Pham Do and Norde. It can be obtained using the average approach by defining $\hat{v}^{free}(S) \stackrel{\text{def}}{=} v(S, \{\{i\} : i \in N_{-S}\} \cup \{S\})$ and $\varphi_i^{free}(v) \stackrel{\text{def}}{=} Sh_i(\hat{v}^{free})$. Hence, the value of $S$ is taken from the partition, in which no externalities from merging coalitions affect it.

The second one, proposed by McQuillin, is dual to Pham Do and Norde. Here, the value of $S$ is taken from the partition, in which all other agents are in one coalition: $\hat{v}^{McQ}(S) \stackrel{\text{def}}{=} v(S, \{N_{-S}, S\})$. As we can see, this value of $S$ is affected by all externalities from merging coalitions. McQuillin value takes the form $\varphi_i^{McQ}(v) \stackrel{\text{def}}{=} Sh_i(\hat{v}^{McQ})$.

*Example 2.2.* Consider a simple, 3-player game:

$$\begin{array}{llll}
\pi_1 = \{\{1\},\{2\},\{3\}\} & v(\{1\},\pi_1) = 4 & v(\{2\},\pi_1) = 2 & v(\{3\},\pi_1) = 3 \\
\pi_2 = \{\{1\},\{2,3\}\} & v(\{1\},\pi_2) = 2 & v(\{2,3\},\pi_2) = 5 & \\
\pi_3 = \{\{1,3\},\{2\}\} & v(\{1,3\},\pi_3) = 7 & v(\{2\},\pi_3) = 0 & \\
\pi_4 = \{\{1,2\},\{3\}\} & v(\{1,2\},\pi_4) = 6 & v(\{3\},\pi_4) = 2 & \\
\pi_5 = \{\{1,2,3\},\emptyset\} & v(\{1,2,3\},\pi_5) = 10 & &
\end{array}$$

Then, games $\hat{v}^{free}$ and $\hat{v}^{McQ}$ are calculated as follows:

$$\begin{array}{ll}
\hat{v}^{free}(\{1\}) = v(\{1\},\pi_1) = 4 & \hat{v}^{McQ}(\{1\}) = v(\{1\},\pi_2) = 2 \\
\hat{v}^{free}(\{2\}) = v(\{2\},\pi_1) = 2 & \hat{v}^{McQ}(\{2\}) = v(\{2\},\pi_3) = 0 \\
\hat{v}^{free}(\{3\}) = v(\{3\},\pi_1) = 3 & \hat{v}^{McQ}(\{3\}) = v(\{3\},\pi_4) = 2 \\
\hat{v}^{free}(\{1,2\}) = v(\{1,2\},\pi_4) = 6 & \hat{v}^{McQ}(\{1,2\}) = v(\{1,2\},\pi_4) = 6 \\
\hat{v}^{free}(\{1,3\}) = v(\{1,3\},\pi_3) = 7 & \hat{v}^{McQ}(\{1,3\}) = v(\{1,3\},\pi_3) = 7 \\
\hat{v}^{free}(\{2,3\}) = v(\{2,3\},\pi_2) = 5 & \hat{v}^{McQ}(\{2,3\}) = v(\{2,3\},\pi_2) = 5 \\
\hat{v}^{free}(\{1,2,3\}) = v(\{1,2,3\},\pi_5) = 10 & \hat{v}^{McQ}(\{1,2,3\}) = v(\{1,2,3\},\pi_5) = 10
\end{array}$$

Based on standard Shapley value: $\varphi^{free} = (\frac{26}{6}, \frac{14}{6}, \frac{20}{6})$ and $\varphi^{McQ} = (\frac{25}{6}, \frac{13}{6}, \frac{22}{6})$.

## 3  Axiomatic Characterization

In this section we will present our axioms including a new definition of the marginal contribution. We will also briefly compare it to the definition proposed by De Clippel and Serrano.

Shapley value is based on four elementary axioms: Efficiency, Symmetry, Additivity and Null-player Axiom. The first three are easily translated to games with externalities.

**Definition 3.1.** *(Efficiency) Function $\varphi$ satisfies Efficiency if the whole payoff is distributed among agents, i.e. $\sum_{i \in N} \varphi_i(v) = v(N, \{N, \emptyset\})$ for every game $v$.*

Let $\sigma : N \to N$ be a permutation of the set of agents. Then
- for every coalition $S \subseteq N$, $\sigma(S) \stackrel{\text{def}}{=} \{\sigma(i) : i \in S\}$
- for every partition $P \in \mathcal{P}$, $\sigma(P) \stackrel{\text{def}}{=} \{\sigma(S) : S \in P\}$ .

The *permutation of game $\sigma(v)$* is a game defined on every embedded coalition by $\sigma(v)(S, P) \stackrel{\text{def}}{=} v(\sigma(S), \sigma(P))$ and the *permutation of $\sigma(\varphi(v))$* is the vector $(\varphi_{\sigma(i)}(v))_{i \in N}$.

**Definition 3.2.** *(Symmetry) Function $\varphi$ satisfies Symmetry if agents' values do not depend on their names, i.e. $\varphi(\sigma(v)) = \sigma(\varphi(v))$ for every game $v$ and every permutation $\sigma$.*

It is widely accepted ([5,4,2]) to translate Additivity as the Linearity in the context of externalities.[1]

**Definition 3.3.** *(Linearity) Function $\varphi$ satisfies Linearity if:*

(a) *for every two games $v_1, v_2$, we have $\varphi(v_1 + v_2) = \varphi(v_1) + \varphi(v_2)$, where $v_1 + v_2$ is a game defined by $(v_1 + v_2)(S, P) = v_1(S, P) + v_2(S, P)$*
(b) *for every game $v$ and constant $\lambda \in \mathbb{R}$, we have $\varphi(\lambda v) = \lambda \cdot \varphi(v)$, where $\lambda v$ is a game defined by $(\lambda v)(S, P) = \lambda \cdot v(S, P)$.*

Our key axiom will be based on the marginality principle. When there are no externalities, the marginal contribution of an agent $i$ to a coalition $S$ can be easily calculated as a difference between the coalition value with and without an agent $i$: $\hat{v}(S) - \hat{v}(S_{-i})$. But when the externalities exist, the value of a coalition $S$ (embedded in $P$) without an agent $i$ depends on where the agent $i$ is. Let us define the *elementary marginal contribution* $mc_{(i,S,P,T)}$ of an agent $i$ to $(S, P)$ in comparison to $i$ being in $T \in P_{-S} \cup \{\emptyset\}$ as a difference between the value of $(S, P)$ and $(S_{-i}, \tau_i^{S,T}(P))$. Then the marginal contribution is the (weighted) average of the elementary marginal contributions:

$$mc_{(i,S,P)}(v) = \sum_{T \in P_{-S} \cup \{\emptyset\}} \alpha_{(i,S,P,T)}(v(S, P) - v(S_{-i}, \tau_i^{S,T}(P)))$$

The empty set in the sum corresponds to the partition in which $i$ is in a singleton coalition $\{i\}$.[2] For our later discussion it would be convenient to consider the elementary marginal contribution $v(S, P) - v(S_{-i}, \tau_i^{S,T}(P))$ as a cost of the agent's $i$ transfer to a coalition $T$.

---

[1] Shapley based his value on Additivity – part (a) of our axiom – as it (combined with his three other axioms) implies part (b) – a very intuitive assumption that when we multiply every value in the game by some scalar, agents' share will increase respectively (i.e. the ratio of agents' share will not change). As shown in [4], the standard Shapley's axioms translated to the games with externalities are too weak to imply full Linearity. Thus, in the presence of externalities, Additivity is usually strengthened to the Linearity.

[2] Note that when $(S, P) = (N, \{N, \emptyset\})$ then $P_{-S} \cup \{\emptyset\} = \{\emptyset\}$, so $mc_{(i,N,\{N,\emptyset\})}(v) = \alpha_{(i,N,\{N,\emptyset\},\emptyset)}(v(N, \{N, \emptyset\}) - v(N_{-i}, \{N_{-i}, \{i\}\}))$.

De Clippel and Serrano used only one non-zero weight for the transfer of $i$ to the empty coalition: $\alpha_{(i,S,P,\emptyset)} = 1$ and $\alpha_{(i,S,P,T)} = 0$ for $T \in P_{-S}$. Their definition of the marginal contribution takes the form:

$$mc^{free}_{(i,S,P)}(v) \stackrel{\text{def}}{=} v(S,P) - v(S_{-i}, P_{-S} \cup \{S_{-i}, \{i\}\})$$

This approach is justified by treating the transfer as a two-step process. In the first step, agent $i$ leaves the coalition $S$ and for a moment remains alone (i.e., creates a singleton coalition). An optional second step consists of agent $i$ joining some coalition from $P_{-S}$ (in coalition terms, $\{i\}$ merges with some other coalition). Although both steps may change the value of $S_{-i}$, the authors argue that only the first one corresponds to the *intrinsic* marginal contribution – the influence from the second step comes rather from the external effect of merging coalitions, not from $i$ leaving $S$. Discarding the impact of merging coalitions in marginal contribution allowed them to derive an "externality-free" value.

We will consider the transfer of $i$ in a different way. Our first step will consist of leaving coalition $S$ and joining one of the other coalitions in partition. In the second step, agent $i$ can exit his new coalition and create his own. Thus, we look at *creating new coalition* as an extra action, which should not be included in the effect of $i$ leaving coalition $S$. According to this, the natural way to define the *steady marginal contribution* of an agent $i$ to $(S,P)$, is to take into account only the transfer to the other existing coalition.

**Definition 3.4.** *The* steady marginal contribution *of an agent $i \in S$ to the embedded coalition $(S,P) \in EC$ is defined as:*

$$mc^{full}_{(i,S,P)}(v) \stackrel{\text{def}}{=} \sum_{T \in P_{-S}} (v(S,P) - v(S_{-i}, \tau_i^{S,T}(P)))$$

*Then, $mc_i^{full}(v) \stackrel{\text{def}}{=} (mc^{full}_{(i,S,P)}(v))_{(S,P) \in EC, i \in S}$ is a vector of steady marginal contributions.*

Our approach can be justified by these real life examples in which creation of a new coalition is rare and not likely. These include political parties or million-dollar industries (such as oil oligopoly). In all such situations, our approach is likely to lead to more proper results.

Based on the definition of steady marginal contribution we can introduce the last axiom, which is our version of the standard Null-player Axiom. In the literature on games with externalities, it is common to assume ([4,5]) that agent $i$ is a null-player when all of his elementary marginal contributions are equal to zero ($v(S,P) - v(S_{-i}, \tau_i^{S,T}(P)) = 0$ for each $(S,P) \in EC$ such that $i \in S$ and $T \in P_{-S} \cup \{\emptyset\}$). Our definition of null-player will differ – we will consider an agent as a null-player when all of his steady marginal contributions are equal to zero.[3]

**Definition 3.5.** *(Null-player Axiom in a steady marginal contribution sense) Function $\varphi$ satisfies Null-player Axiom if for every agent $i$ such that vector of steady marginal contributions $mc_i^{full}(v)$ is a zero vector occurs $\varphi_i(v) = 0$.*

---

[3] It is easy to prove that our axiom strengthens the standard one, as the fact that all of the agent's $i$ elementary marginal contributions equal zero implies that all his steady marginal contributions are equal to zero as well.

## 4 Constant-Coalition Games

In this section we will introduce a new class of simple games – *constant-coalition games*. This games will play a key role in a proof of the uniqueness of the value in the next section. The name comes from the fact that, in a given game, every partition in which a coalition with non-zero value is embedded, has exactly the same number of coalitions. We show that the collection of such games is a basis of partition function games.

First, we will need some additional notation.

**Definition 4.1.** *($R_1 \preceq R_2$) Let $R_1, R_2$ be two proper, non-empty subsets of two partitions. We say that $R_2$ can be reduced to $R_1$ (denoted $R_1 \preceq R_2$) if three conditions are met:*

(a) *all agents which appear in $R_1$, appear in $R_2$ (i.e. $\bigcup_{T_1 \in R_1} T_1 \subseteq \bigcup_{T_2 \in R_2} T_2$)*
(b) *two agents which are in the same coalition in $R_1$, are in the same coalition in $R_2$*
(c) *two agents which are not in the same coalition in $R_1$ are not in the same coalition in $R_2$.*

Assume $R_1 \preceq R_2$. Based on the presented conditions, as we delete agents from $R_2$ which are not in $R_1$, we get exactly the $R_1$ configuration. This observation can be expressed in an alternative definition of the $\preceq$-operator.

**Proposition 4.2.** *Let $R_1, R_2$ be two proper, non-empty subsets of two partitions. Then:*[4]

$$R_1 \preceq R_2 \Leftrightarrow \exists_{S \subseteq N} R_1 = \{T_2 \setminus S : T_2 \in R_2 \text{ and } T_2 \not\subseteq S\} \vee R_1 = \{\emptyset\}$$

For example $\{\{1,2\},\{3\}\} \preceq \{\{1,2,4\},\{3\},\{5\}\}$ but $\{\{1,2\},\{3\}\} \not\preceq \{\{1,2,3\}\}$ and $\{\{1,2\},\{3\}\} \not\preceq \{\{1\},\{3,4\}\}$.

Now we can introduce our new basis for games with externalities.

**Definition 4.3.** *For every embedded coalition $(S, P)$, the* constant-coalition game $e^{(S,P)}$ *is defined by*

$$e^{(S,P)}(\tilde{S}, \tilde{P}) \stackrel{\text{def}}{=} \begin{cases} (|P| - 1)^{-(|\tilde{S}\setminus S|)} & \text{if } (\tilde{P}_{-\tilde{S}} \preceq P_{-S}) \text{ and } (|P| = |\tilde{P}|), \\ 0 & \text{otherwise,} \end{cases}$$

*for every $(\tilde{S}, \tilde{P}) \in EC$.*

Note that $(\tilde{P}_{-\tilde{S}} \preceq P_{-S})$ implies $S \subseteq \tilde{S}$ as we get $\tilde{N} \setminus \tilde{S} \subseteq N \setminus S$ from the (a) condition in $\preceq$-operator definition. So, in our game, $e^{(S,P)}$ non-zero values have only embedded coalitions formed from $(S, P)$ by some transition of agents from $P \setminus \{S\}$ to $S$ which does not change the number of the coalitions.

**Lemma 4.4.** *The collection of constant-coalition games is a basis of the partition function games.*

---

[4] The equivalence of the definitions when $R_1$ and $R_2$ contain only non-empty coalitions is easy to see. As the only partition which contains an empty set is $\{N, \emptyset\}$ then the only proper, non-empty subset of the partition which contains an empty set is $\{\emptyset\}$. Based on the first definition $\{\emptyset\} \preceq R_2$ for every $R_2$ and $R_1 \preceq \{\emptyset\}$ implies $R_1 = \{\emptyset\}$ (as we don't allow empty $R_1$).

*Proof.* Let $e = (e^{(S,P)})_{(S,P) \in EC}$ be the vector of all games.

First, we will show that the constant-coalition games are linearly independent. Suppose the contrary. Then there exists a vector of weights $\alpha = (\alpha_{(S,P)})_{(S,P) \in EC}$ with at least one non-zero value such that $\alpha \times e = \sum_{(S,P) \in EC} \alpha_{(S,P)} e^{(S,P)}$ is a zero vector. Let $(S^*, P^*)$ be the embedded coalition with a non-zero weight $\alpha_{(S^*,P^*)} \neq 0$ and minimal $S^*$ (i.e. $(S^*, P^*)$ is the minimal element of the embedded-coalition relation $r$: $(S_1, P_1) r (S_2, P_2) \Leftrightarrow S_1 \subseteq S_2$). So, for any other game $e^{(S,P)}$ either $\alpha_{(S,P)} = 0$ or $S \not\subseteq S^* \Rightarrow e^{(S,P)}(S^*, P^*) = 0$ (the implication follows from the remarks after Definition 4.3). Then

$$\sum_{(S,P) \in EC} \alpha_{(S,P)} e^{(S,P)}(S^*, P^*) = \alpha_{(S^*,P^*)} e^{(S^*,P^*)}(S^*, P^*) = \alpha_{(S^*,P^*)} \neq 0,$$

contrary to the previous assumption.

The size of a collection of all the constant-coalition games is equal to the dimension of the partition function games space, hence the collection must be a basis. □

## 5 Uniqueness of the Value

In this section we show that there exists only one value that satisfies all the introduced axioms and that it is equivalent to the value proposed by McQuillin.

**Theorem 5.1.** *There is a unique value $\varphi^{full}$ satisfying Efficiency, Symmetry, Linearity and Null-player Axiom (in a steady marginality sense).*

*Proof.* We will show that in every coalition-constant game there exists only one value which satisfies those axioms. Based on Linearity and Lemma 4.4 this will imply our theorem.

Let $e^{(S,P)}$ be one of the coalition-constant games. We will show that any player $i$ from the coalition other than $S$ is a null-player (in a steady marginality sense). Based on the Definition 3.5 we have to prove that $mc_i(e^{(S,P)})$ is a zero vector. So, for every $(\tilde{S}, \tilde{P}) \in EC$ such that $i \in \tilde{S}$:

$$mc_{(i,\tilde{S},\tilde{P})}(e^{(S,P)}) = \sum_{\tilde{T} \in \tilde{P} \setminus \{\tilde{S}\}} e^{(S,P)}(\tilde{S}, \tilde{P}) - e^{(S,P)}(\tilde{S}_{-i}, \tau_i^{\tilde{S},\tilde{T}}(\tilde{P})) = 0$$

We divide the proof into two cases with zero and non-zero value of $e^{(S,P)}(\tilde{S}, \tilde{P})$.

**Lemma 5.2.** *If $e^{(S,P)}(\tilde{S}, \tilde{P}) = 0$ then, for every $T \in \tilde{P} \setminus \{\tilde{S}\}$,*

$$e^{(S,P)}(\tilde{S}_{-i}, \tau_i^{\tilde{S},T}(\tilde{P})) = 0.$$

*Proof.* Based on the definition of the coalition-constant games we can deduce that at least one of the following conditions occurs:

- $\tilde{P}_{-\tilde{S}} \not\preceq P_{-S}$ - from the definition of $\preceq$-operator we know that there is an agent in $\tilde{P}_{-\tilde{S}}$ which is not in $P_{-S}$, or there is a pair of agents which are together in one and not together in the other structure; it is easy to see, that adding player $i$ to some coalition in $\tilde{P}_{-\tilde{S}}$ will not fix any of these anomalies;

– $|P| \neq |\tilde{P}|$ - if $i$ is alone ($\tilde{S} = \{i\}$) then, for every $T \in \tilde{P} \setminus \{\tilde{S}\}$,

$$e^{(S,P)}(\tilde{S}_{-i}, \tau_i^{\tilde{S},T}(\tilde{P})) = e^{(S,P)}(\emptyset, \tau_i^{\tilde{S},T}(\tilde{P})) = 0;$$

otherwise, as we only consider the transfer of an agent $i$ to the other existing coalition, the number of the coalitions remains intact: $|P| \neq |\tau_i^{\tilde{S},T}(\tilde{P})|$. □

**Lemma 5.3.** *If $e^{(S,P)}(\tilde{S}, \tilde{P}) = x$ for $x > 0$ then there exists only one $T^* \in \tilde{P} \setminus \{\tilde{S}\}$ such that $e^{(S,P)}(\tilde{S}_{-i}, \tau_i^{\tilde{S},T^*}(\tilde{P}))$ has non-zero value. Moreover, this value is equal to $x(|P| - 1)$.*

*Proof.* If $e^{(S,P)}(\tilde{S}, \tilde{P}) > 0$ and $e^{(S,P)}(\tilde{S}_{-i}, \tau_i^{\tilde{S},T}(\tilde{P})) > 0$ then from the definition:

$$e^{(S,P)}(\tilde{S}_{-i}, \tau_i^{\tilde{S},T}(\tilde{P})) = (|P| - 1)^{-|\tilde{S}_{-i} \setminus S|} = (|P| - 1) \cdot (|P| - 1)^{-|\tilde{S} \setminus S|}$$
$$= (|P| - 1) \cdot e^{(S,P)}(\tilde{S}, \tilde{P})$$

which proves the values equality part.

First we will consider a special case when $\tilde{P}$ contains the empty coalition. Then $\tilde{P} = \{N, \emptyset\}$ (as it is the only partition with the empty coalition) and the only transition of $i$ is allowed to the empty coalition: $(\tilde{S}, \tau_i^{\tilde{S},T}(\tilde{P}))$ is equal $(N_{-i}, \{N_{-i}, \{i\}\})$. As we assumed $i \notin S$, we know that $\{i\} \preceq P_{-S}$ and as we're not changing the partition size $(|\{N, \emptyset\}| = |\{N_{-i}, \{i\}\}|)$ this follows that $e^{(S,P)}(N_{-i}, \{N_{-i}, \{i\}\}) > 0$ and finishes this case.

Now let's assume that all the coalitions in $\tilde{P}$ are non-empty. Let $T_i \in P \setminus S$ be the agent's $i$ coalition. From the constant-games definition we know that $P_{-S}$ can be reduced to $\tilde{P}_{-\tilde{S}}$ and that in both there is the same number of the coalitions: $|P_{-S}| = |\tilde{P}_{-\tilde{S}}|$. As the agents from one coalition cannot be separated, there must be some non-empty coalition $\tilde{T}_i$ in $\tilde{P}_{-\tilde{S}}$ which can be reduced from $T_i$ by deleting agents from $\tilde{S}$. It must contain at least one agent denoted by $j$ (and $j \neq i$, because $i \in \tilde{S}$). So when we consider a transition to any other coalition than $\tilde{T}_i$ we will separate $i$ and $j$ agents which will violate (c) condition in $\preceq$-operator definition and imply zero value in $e^{(S,P)}$ game. But in $\tau_i^{\tilde{S},\tilde{T}_i}(P)$ all the conditions will be satisfied – (a) is obviously satisfied as $i \notin S$ and $\tilde{P}_{-\tilde{S}}$ was already a subset of $P_{-S}$; (b) and (c) are satisfied because the relations between an additional $i$ agent are equal to the relations of $j$ who is already in the structure.

Again, we do not change the size of the partition. We have to check only one special case when $\tilde{S} = \{i\}$. But from $e^{(S,P)}(\tilde{S}, \tilde{P}) > 0$ we get $S \subseteq \tilde{S}$ and as we know that $i \notin S$ we get $S = \emptyset$ which means that the game $e^{(S,P)}$ is incorrect.

So, finally: $e^{(S,P)}(\tilde{S}_{-i}, \tau_i^{\tilde{S},\tilde{T}_i}(\tilde{P})) > 0$. □

From Lemma 5.2 and Lemma 5.3 we have that every agent $i \notin S$ is a null-player (in a steady marginality sense). Based on our Null-player Axiom (Definition 3.5), $\varphi_i^{full}(e^{(S,P)}) = 0$ and based on Symmetry (Definition 3.2) and Efficiency (Definition 3.1) we get:

$$\varphi_j^{full}(e^{(S,P)}) = \frac{1}{|S|} \cdot \sum_{j \in S} \varphi_j^{full}(e^{(S,P)}) = \frac{1}{|S|} \cdot e^{(S,P)}(N, \{N, \emptyset\}).$$

As our value $\varphi^{full}$ clearly satisfies Efficiency, Linearity and Symmetry, the only observation we need to add is that it also satisfies Null-player Axiom. As agents not from $S$ are null-players and get nothing it would be sufficient to show that no agent from $S$ is a null-player. But every agent $j$ from $S$ has a non-zero marginal contribution to $(S, P)$: $e^{(S,P)}(S, P) = 1$ and $e^{(S,P)}(S_{-j}, \tau_j^{S,T}(P)) = 0$ for every $T \in P \setminus S$. That finishes the proof of Theorem 5.1. □

Now let us show that our unique value is indeed equal to the value proposed by McQuillin.

**Theorem 5.4.** *Let $v$ be a game with externalities. Then $\varphi^{McQ}(v) = \varphi^{full}(v)$.*

*Proof.* Again, based on Linearity, we will show the adequacy on the constant-coalition games. In the proof of Theorem 5.1 we have showed that $\varphi_i^{full}(e^{(S,P)}) = 0$ for every $i \notin S$ and $\varphi_j^{full}(e^{(S,P)}) = \frac{1}{|S|} \cdot e^{(S,P)}(N, \{N, \emptyset\})$ for every $j \in S$.

Let $(S, P)$ be an embedded coalition. Assume that $|P| > 2$. As $|\{N, \emptyset\}| = 2 \neq |P|$, based on the definition of the constant-coalition games (Definition 4.3) we get $e^{(S,P)}(N, \{N, \emptyset\}) = 0$ and $\varphi_i^{full}(e^{(S,P)}) = 0$ for every agent $i \in N$. Also $\varphi_i^{McQ}(e^{(S,P)}) = 0$ for every agent $i \in N$, because $\hat{v}^{McQ}(S) = 0$ for every $S \subseteq N$ as no embedded coalition of form $(\tilde{S}, \{\tilde{S}, N \setminus \tilde{S}\})$ has a non-zero value in $e^{(S,P)}$ (the reason here is the same – the partitions sizes do not match).

If $|P| = 2$ then embedded coalition has the form $(S, \{S, N \setminus S\})$ and $e^{(S, \{S, N \setminus S\})}$ assigns a non-zero value (equal 1) only to an embedded coalition $(\tilde{S}, \{\tilde{S}, N \setminus \tilde{S}\})$ such that $S \subseteq \tilde{S}$. Hence, $\hat{v}^{McQ}(\tilde{S}) = 1$ when $S \subseteq \tilde{S}$ and $\hat{v}(\tilde{S}) = 0$ otherwise. Based on the basic Shapley's axioms for $\hat{v}^{McQ}$ we get that $\varphi_i^{McQ}(e^{(S,P)}) = 0$ for $i \notin S$ and $\varphi_j^{McQ}(e^{(S,P)}) = Sh_j(\hat{v}) = \frac{1}{|S|}$ for $j \in S$.

Let's check if our value has the same results. As mentioned at the beginning of the proof, for $i \notin S$, $\varphi_i^{full}(e^{(S,P)}) = 0$ and for $j \in S$, $\varphi_j^{full}(e^{(S,P)}) = \frac{1}{|S|} \cdot e^{(S,P)}(N, \{N, \emptyset\}) = \frac{1}{|S|}$ which completes the proof. □

## 6   A Comparison of Various Marginality Definitions

In this section we examine various approaches to marginality more broadly and we compare them with steady marginality.

In Section 3 we have presented the universal definition of marginality:

$$mc_{(i,S,P)}(v) = \sum_{T \in P_{-S} \cup \{\emptyset\}} \alpha_{(i,S,P,T)}(v(S,P) - v(S_{-i}, \tau_i^{S,T}(P)))$$

Based on the accepted definition of marginal contribution we define a vector of marginal contributions $mc_i(v) = (mc_{(i,S,P)}(v))_{(S,P) \in EC, i \in S}$.

It seems reasonable to normalize the weights by assuming that the sum of them is equal 1: $\sum_{T \in P_{-S} \cup \{\emptyset\}} \alpha_{(i,S,P,T)} = 1$ for each $(S, P) \in EC$ such that $i \in S$. As we have considered only 0-1 weights we have omitted this step to increase the clarity of the presentation. But it is important to notice, that with respect to the conventional axioms based on the marginality this normalization is not significant as we compare $mc_{(i,S,P)}$ only to the same marginal contribution in other game (in Bolger [5] and De Clipper and Serrano [7] from $mc_i(v_1) = mc_i(v_2)$ we conclude $\varphi_i(v_1) = \varphi_i(v_2)$) or to zero (in Hu and Yang [6] and our paper from $mc_i(v_1) = 0$ we conclude that $\varphi_i(v) = 0$). Thus, the only important aspect is the weight ratio.

In all the definitions of marginal contribution proposed in the literature $\alpha_{(i,N,\{N,\emptyset\},\emptyset)} = 1$.[5] Thus, in the rest of this section we assume, that $P \neq \{N, \emptyset\}$.

Chronologically, the first definition of marginality proposed is also the most intuitive one. Bolger [5] defined the marginality as a simple average of all the elementary marginal contributions:

$$\alpha^B_{(i,S,P,T)} = 1 \text{ for } T \in P_{-S} \cup \{\emptyset\}$$

In his paper, Bolger studied the *basic games* – games with only 0-1 values. Thus, his marginal contribution is the number of partitions from which agent's transfer turns the $S_{-i}$ coalition value (negative if value of $S$ is zero). Unfortunately, there is no closed form expression for Bolger's value.

Steady marginality is quite similar to Bolger's one:

$$\alpha^{full}_{(i,S,P,T)} = 1 \text{ for } T \in P_{-S} \quad \text{and} \quad \alpha^{full}_{(i,S,P,T)} = 0 \text{ otherwise}$$

We have also already introduced the marginality proposed by De Clipper and Serrano:

$$\alpha^{free}_{(i,S,P,T)} = 1 \text{ for } T = \emptyset \quad \text{and} \quad \alpha^{free}_{(i,S,P,T)} = 0 \text{ otherwise}$$

It is easy to see, that our marginality complements De Clipper and Serrano's marginality to Bolger's one.

Another approach was proposed by Hu and Yang [6]. Their marginality assigns the same weight to the transfer to every existing coalition and *higher* value for the partition where $i$ is alone:

$$\alpha^{HY}_{(i,S,P,T)} = 1 \text{ for } T \in P_{-S} \quad \text{and} \quad \alpha^{HY}_{(i,S,P,T)} = 1 + r \text{ otherwise}$$

where $1 + r = \frac{|\{P^* \in \mathcal{P} : (\tau_i^{S,\emptyset}(P) \setminus \{S_{-i}\}) \preceq P^*\}|}{|\{P^* \in \mathcal{P} : (\tau_i^{S,T}(P) \setminus \{S_{-i}\}) \preceq P^*\}|}$ for any $T \in P_{-S}$.[6] It can be shown, that $r \geq 0$.

Let us consider a simple environment in which there exists $m$ political parties – $S^{(j)}$ for every $j \in \{1, 2, \ldots, m\}$ – and one independent agent $i$ (i.e. $N = \{i\} \cup \bigcup_j S^{(j)}$). Let $P = \{S^{(j)} : j \in \{1, 2, \ldots, m\}\} \cup \{\{i\}\}$ and define $P^{(j)} = \tau_i^{\{i\}, S^{(j)}}(P)$. Assume that an agent $i$ does not have any political power on his own, but by joining one of the parties he increases the coalition value by $(m-1)$ and decreases the values of other parties by 1, thus: $v(S^{(j)}_{+i}, P^{(j)}) = v(S^{(j)}, P) + (m-1)$ and $v(S^{(j)}, P^{(k)}) = v(S^{(j)}, P) - 1$ for every $S^{(j)}, S^{(k)} \in P$ and $S^{(j)} \neq S^{(k)}$.

Consider the marginal contribution of an agent $i$ to the coalition $S^{(j)}_{+i}$. As we normalize weights (in a way mentioned before) based on non-existing marginality we get $mc^{full}_{(i,S^{(j)}_{+i},P^{(j)})} = m$ as we exclude a non-realistic situation from the consideration – when agent $i$ decides to waste his potential. De Clippel and Serrano consider only this one situation and get $mc^{free}_{(i,S^{(j)}_{+i},P^{(j)})} = m - 1$. Bolger reaches some compromise, as he does not differentiate partitions: $mc^{B}_{(i,S^{(j)}_{+i},P^{(j)})} = m - \frac{1}{m}$. The last marginal contribution $mc^{HY}_{(i,S^{(j)}_{+i},P^{(j)})}$ is slightly smaller than Bolger's and depends on the size of $S^{(j)}$ and the number of parties.

---

[5] Assume that $\alpha_{(i,N,\{N,\emptyset\},\emptyset)} = 0$, then $mc_{(i,N,\{N,\emptyset\})} = 0$ regardless of $v(N, \{N,\emptyset\})$ and $v(N_{-i}, \{N, \{i\}\})$. So, in game with only one non-zero value for grand-coalition $v(N, \{N, \emptyset\}) = 1$ all the marginal contributions are equal zero which is unintuitive and with every axiom based on marginality results in a contradiction.

[6] That definition may seem wrong, as it might not be obvious, why $|\{P^* \in \mathcal{P} : (\tau_i^{S,T}(P) \setminus \{S_{-i}\}) \preceq P^*\}|$ is equal for every $T \in P_{-S}$. It appears, that $|\{P^* \in \mathcal{P} : \tilde{P}_{-\tilde{S}} \preceq P^*\}|$ depends only on $|\tilde{P}|$ and $|\tilde{S}|$ and grows with increasing $|\tilde{P}|$. That also explains why the numerator is larger than the denominator in $1 + r$ fractional definition.

## 7  Conclusions

In this paper we have studied the problem of finding a fair division of jointly gained payoff in coalition games with externalities. We have presented an innovative approach to marginality in which the contribution of an agent to the coalition is evaluated only in reference to the partitions where the agent is not alone. This allowed us to define a new version of Null-player Axiom which, together with Efficiency, Symmetry and Linearity, uniquely determines a division scheme. We have proved that this value is equal to the one proposed earlier by McQuillin.

Our work can be extended in various directions. It is not clear if any of the adopted axioms can be dropped (just as De Clippel and Serrano based their value only on Symmetry, Efficiency and their version of Marginality axiom). Another question is whether there exist any other definitions of marginality which lead to McQuillin's value. Looking a bit further, it would be interesting to find any universal link between the definition of marginality and the formula for the value derived from it. Finally, other approaches to marginality axioms can be studied.